# Analyzing the Motion of a Charged Rigid Body under the Influence of a Gyrostatic Torque

A. H. Elneklawy
Department of Mathematics, Faculty of Science, Kafrelsheikh University, Kafr El-Sheikh 33516, Egypt
ahmed_hassan@sci.kfs.edu.eg

*Abstract*: This research investigates the rotational dynamics of a charged axisymmetric spinning rigid body influenced by gyrostatic torque. The study also accounts for the effects of transverse and constant body-fixed torques and an electromagnetic force field. Euler's equations of motion are employed to formulate the governing equations for the system. Given the absence of torque along the spin axis and the nearly symmetric structure of the rigid body, the spin rate remains almost constant. By assuming small angular deviations of the spin axis from a fixed spatial direction, approximate analytical solutions in closed form are derived for the attitude, translational, and rotational motions. These solutions, expressed in a compact complex form, provide an effective tool for analyzing the maneuvers of rigid bodies spinning. The study derives analytical solutions for several variables, including angular velocities, Euler angles, transverse and axial displacements, and velocities. Graphical simulations of the solutions demonstrate their accuracy, while additional graphs illustrate the positive effects of varying body parameters on the motion. This work is significant in diverse scientific and engineering fields, offering insights that enhance the design of mechanical systems, explain celestial dynamics, and improve spacecraft performance.

## I. Introduction

In the past century, there has been a significant amount of research focused on addressing Euler's equations for the rigid body's rotational motion. This involves studying the forces and torques acting on the body, applying rotational dynamics principles, and accurately calculating relevant quantities and parameters to describe the body's rotation. Numerous researchers have been attentive to this topic, as shown in [1-11]. The governing equations of motion for such a problem has been approached in all of the aforementioned references especially in [1-2]. The formulations for the equations of Euler's angles with their different sequences have been presented in [3]. In [4], the author obtained an estimated analytic solution for the rotary motion of a nearly symmetrical rigid body that experiences constant body-fixed torques. The approximate solution for Euler's equations of motion is precisely expressed using Fresnel integrals [5] and is accurate for bodies with symmetry. A rough resolution for the Euler's angles is determined with Fresnel integrals, as well as sine and cosine integrals. An analysis of the influence of gyrostatic torque, constant body-fixed torques, and resistant forces on the motion of a charged rigid body is given in [6]. A suitable governing system for equations of motion is approached using the averaging method. To reach the required results, Taylor's method is used along with some initial conditions to solve the averaged equations of motion of the rigid body. Various values related to gyrostatic torque, electromagnetic force field, constant body-fixed torques, and torque of a resistive force are illustrated with diagrams. It has also been discussed whether the motion of the rigid body is stable or not. In [7-11], the author's goal was to investigate the rotational movement of an asymmetric rigid body subjected to constant body-fixed torques and a non-zero first component gyrostatic torque. They employed Euler's equations of motion to derive a series of dimensionless equations, which were then suggested for analyzing the stability of equilibrium points.

The examination of the impact of gyrostatic torque on the rotary motion of a spinning axisymmetric charged rigid body is provided in this study. Additionally, the influence of transverse and constant body-fixed torques and electromagnetic force field are considered. Euler's equation is used to derive the equations of motion that governs the rigid body's rotation. Due to the lack of torque exerted along the spin axis and the rigid body possesses a nearly symmetrical structure, the spin rate remains almost constant. Assuming small angular deviations of the spin axis regarding a stationary direction in space, it is possible to obtain approximate analytical solutions in closed form for the attitude, translational, and rotational movements. These concise solutions, expressed in complex form, are highly effective in analyzing the maneuvers performed by spinning rigid bodies. The approach focuses on deriving the analytical solutions for various variables including angular velocities, Euler's angles, angular momentum, transverse velocities, transverse displacements, axial velocity, and axial displacement. To demonstrate the accuracy of the method used, graphical simulations of these solutions are presented.

## II. The problem's articulating

Therefore, Euler's equations of motion governing the rigid body's motion are given by [1,2]

$$D_1\dot{p} + (D_3 - D_2)qr + (\lambda_3 - eHl^2\cos\sigma)q = M_1,$$
$$D_2\dot{q} + (D_1 - D_3)rp - (\lambda_3 - eHl^2\cos\sigma)p = M_2, \qquad (1)$$
$$D_3\dot{r} + (D_2 - D_1)pq = M_3.$$

where $M_j\ (j = 1, 2, 3)$ are the constant body-fixed torques, and the dot represents the differentiation concerning time $t$.

## III. The analytical solutions to some parameters controlling the problem

This section presents a detailed analytical solution for some parameters that control the motion of the body as follows:

## A. The angular velocities solutions

Considering the assumption that the rigid body is rotating around its $z_2$-axis, it can be inferred that $M_3$ will equal zero. In the case of an axisymmetric rigid body, where $D_1$ is equal to $D_2$, or in a situation where the difference between $D_1$ and $D_2$ is negligible, the solution of system (1) can be expressed as

$$r \approx r_0 = r(0). \tag{2}$$

$$\Pi_1(t) = \Pi_{10} e^{i\zeta t[r_0 + \{(\lambda_3 - eHl^2 \cos\sigma)/\zeta_1 D_1\}]} + \frac{iA}{\zeta[r_0 + \{(\lambda_3 - eHl^2 \cos\sigma)/\zeta_1 D_1\}]} \tag{3}$$
$$\times [1 - e^{i\zeta t[r_0 + \{(\lambda_3 - eHl^2 \cos\sigma)/\zeta_1 D_1\}]}].$$

Comparing the real and imaginary terms on both sides of equation (3) gives

$$p_1 = \text{Re}[\Pi_1(t)], \qquad q_1 = \text{Im}[\Pi_1(t)]. \tag{4}$$

## B. Eulerian angles solutions

The 3-1-2 Euler angle sequence [3], consisting of Euler's angles $\psi, \phi,$ and $\theta$, is used to describe the process of elucidating the correspondence between the reference frame attached to the body and the reference frame that remains fixed in space, as seen in Fig. (1). The associated kinematic equations are as follows

$$\dot{\psi} = p\cos\phi + r\sin\phi,$$
$$\dot{\phi} = q - (r\cos\phi - p\sin\phi)\tan\psi, \tag{5}$$
$$\dot{\theta} = (r\cos\phi - p\sin\phi)\sec\psi.$$

The equations at hand are extremely complex and appear to be difficult to solve. However, significant advancements have been achieved by employing linearization techniques, such as assuming that the variables $\psi$ and $\phi$ are small.

As a result of these simplifications, system (5) is simplified to

$$\dot{\psi} \approx p + r\phi,$$
$$\dot{\phi} \approx q - r\psi, \tag{6}$$
$$\dot{\theta} \approx r.$$

The solutions of (6) are

$$\psi(t) = \text{Re}[\Gamma(t)],$$
$$\phi(t) = \text{Im}[\Gamma(t)], \tag{7}$$
$$\theta(t) = r_0 t + \theta_0.$$

$$\Gamma(t) = \Gamma_0 e^{-ir_0 t} + e^{-ir_0 t} I_\Gamma(t). \tag{8}$$

One can write $I_\Gamma(t)$ in the form

$$I_\Gamma(t) = [(\sqrt{\zeta_1} + \sqrt{\zeta_2})I_{1\Gamma}(t) + (\sqrt{\zeta_1} - \sqrt{\zeta_2})I_{2\Gamma}(t)]/2\zeta, \tag{9}$$

where

$$I_{1\Gamma}(t) = \int_0^t e^{ir_0 \tau} \Pi_1(\tau) d\tau$$
$$= \frac{-i\Pi_{10}}{\zeta[r_0 + \{(\lambda_3 - eHl^2 \cos\sigma)/\zeta_1 D_1\}] + r_0}[e^{it\{\zeta[r_0 + \{(\lambda_3 - eHl^2 \cos\sigma)/\zeta_1 D_1\}] + r_0\}} - 1]$$
$$+ \frac{A}{\zeta[r_0 + \{(\lambda_3 - eHl^2 \cos\sigma)/\zeta_1 D_1\}]}\left\{\frac{e^{ir_0 t} - 1}{r_0} - \frac{1}{\zeta[r_0 + \{(\lambda_3 - eHl^2 \cos\sigma)/\zeta_1 D_1\}] + r_0}\right. \tag{10}$$
$$\left.\times [e^{it\{\zeta[r_0 + \{(\lambda_3 - eHl^2 \cos\sigma)/\zeta_1 D_1\}] + r_0\}} - 1]\right\},$$

$$I_{2\Gamma}(t) = \int_0^t e^{ir_0\tau}\tilde{\Pi}_1(\tau)d\tau$$

$$= \frac{-i\tilde{\Pi}_{10}}{r_0 - \zeta[r_0 + \{(\lambda_3 - eHl^2\cos\sigma)/\zeta_1 D_1\}]}[e^{it\{r_0 - \zeta[r_0 + \{(\lambda_3 - eHl^2\cos\sigma)/\zeta_1 D_1\}]\}} - 1]$$

$$- \frac{\tilde{A}}{\zeta[r_0 + \{(\lambda_3 - eHl^2\cos\sigma)/\zeta_1 D_1\}]}\{\frac{e^{ir_0 t} - 1}{r_0} - \frac{1}{r_0 - \zeta[r_0 + \{(\lambda_3 - eHl^2\cos\sigma)/\zeta_1 D_1\}]}$$

$$\times [e^{it\{r_0 - \zeta[r_0 + \{(\lambda_3 - eHl^2\cos\sigma)/\zeta_1 D_1\}]\}} - 1]\}. \tag{11}$$

## C. Transverse and axial velocities solutions

In the subsequent analysis, we assume that the force related to the inertial moments of the body has the components $f_1, f_2$, and $f_3$ are constants as the applied torques to the body were. The relation between the accelerations of inertia and the body analogous to Newton's second law in the rotational kinematics is expressed as follows [4]

$$\begin{bmatrix}\dot{\omega}_1\\\dot{\omega}_2\\\dot{\omega}_3\end{bmatrix} = \begin{bmatrix}\cos\theta & -\sin\theta & \phi\cos\theta + \psi\sin\theta\\ \sin\theta & \cos\theta & \phi\cos\theta - \psi\cos\theta\\ -\phi & \psi & 1\end{bmatrix}\begin{bmatrix}f_1\\f_2\\f_3\end{bmatrix}, \tag{12}$$

Substituting (4) and (7) for the values of $\theta$ and $\Gamma$ into equation (12) to yield

$$\omega(t) = \omega(0) + ie^{i\theta_0}m^{-1}\{f\,r_0^{-1}(1 - e^{ir_0 t}) - f_3[\Gamma_0 t + I_\omega(t)]\}, \tag{13}$$

where

$$I_\omega(t) = [(\sqrt{\zeta_1} + \sqrt{\zeta_2})I_{1\omega}(t) + (\sqrt{\zeta_1} - \sqrt{\zeta_2})I_{2\omega}(t)]/2\zeta, \tag{14}$$

$$I_{1\omega}(t) = \int_0^t I_{1\Gamma}(\tau)d\tau$$

$$= \frac{-i\tilde{\Pi}_{10}}{\zeta[r_0 + \{(\lambda_3 - eHl^2\cos\sigma)/\zeta_1 D_1\}] + r_0}[\frac{i(1 - e^{it\{\zeta[r_0 + \{(\lambda_3 - eHl^2\cos\sigma)/\zeta_1 D_1\}] + r_0\}})}{\zeta[r_0 + \{(\lambda_3 - eHl^2\cos\sigma)/\zeta_1 D_1\}] + r_0} - t]$$

$$+ \frac{A}{\zeta[r_0 + \{(\lambda_3 - eHl^2\cos\sigma)/\zeta_1 D_1\}]}\{r_0^{-1}[\{ir_0^{-1}(1 - e^{ir_0 t}) - t\} \tag{15}$$

$$- \frac{1}{\zeta[r_0 + \{(\lambda_3 - eHl^2\cos\sigma)/\zeta_1 D_1\}] + r_0}[\frac{i(1 - e^{it\{\zeta[r_0 + \{(\lambda_3 - eHl^2\cos\sigma)/\zeta_1 D_1\}] + r_0\}})}{\zeta[r_0 + \{(\lambda_3 - eHl^2\cos\sigma)/\zeta_1 D_1\}] + r_0} - t]\},$$

$$I_{2\omega}(t) = \int_0^t I_{2\Gamma}(\tau)d\tau$$

$$= \frac{-i\tilde{\Pi}_{10}}{r_0 - \zeta[r_0 + \{(\lambda_3 - eHl^2\cos\sigma)/\zeta_1 D_1\}]}[\frac{i(e^{it\{r_0 - \zeta[r_0 + \{(\lambda_3 - eHl^2\cos\sigma)/\zeta_1 D_1\}]\}} - 1)}{r_0 - \zeta[r_0 + \{(\lambda_3 - eHl^2\cos\sigma)/\zeta_1 D_1\}]} - t]$$

$$- \frac{\tilde{A}}{\zeta[r_0 + \{(\lambda_3 - eHl^2\cos\sigma)/\zeta_1 D_1\}]}\{r_0^{-1}[ir_0^{-1}(1 - e^{ir_0 t}) - t] \tag{16}$$

$$- \frac{1}{r_0 - \zeta[r_0 + \{(\lambda_3 - eHl^2\cos\sigma)/\zeta_1 D_1\}]}[\frac{i(e^{it\{r_0 - \zeta[r_0 + \{(\lambda_3 - eHl^2\cos\sigma)/\zeta_1 D_1\}]\}} - 1)}{r_0 - \zeta[r_0 + \{(\lambda_3 - eHl^2\cos\sigma)/\zeta_1 D_1\}]} - t]\}.$$

Now, we obtain the axial velocity's solution. To achieve this objective, the integration of the third equation in (12) produces this solution, as follows

$$\omega_3(t) = \omega_3(0) + m^{-1}f_3 t + \frac{i}{2m}[\tilde{f}\int_o^t \Gamma(\tau)d\tau - f\int_o^t \tilde{\Gamma}(\tau)d\tau], \tag{17}$$

where

$$\int_0^t \Gamma(\tau)d\tau = \int_0^t [\Gamma_0 e^{-ir_0\tau} + e^{-ir_0\tau}I_\Gamma(\tau)]d\tau$$

$$= i\Gamma_0 r_0^{-1}(e^{-ir_0 t} - 1) + I_{3\omega}(t), \tag{18}$$

and

$$I_{3\omega}(t) = [(\sqrt{\zeta_1} + \sqrt{\zeta_2})I_{4\omega}(t) + (\sqrt{\zeta_1} - \sqrt{\zeta_2})I_{5\omega}(t)]/2\zeta, \tag{19}$$

$$I_{4\omega}(t) = \frac{-i\Pi_{10}}{\zeta[r_0 + \{(\lambda_3 - eHl^2\cos\sigma)/\zeta_1 D_1\}] + r_0} \left\{ \frac{i[e^{it\{\zeta[r_0 + \{(\lambda_3 - eHl^2\cos\sigma)/\zeta_1 D_1\}] + r_0\}} - 1]}{\zeta[r_0 + \{(\lambda_3 - eHl^2\cos\sigma)/\zeta_1 D_1\}] + r_0} - t \right\}$$

$$+ \frac{A}{\zeta[r_0 + \{(\lambda_3 - eHl^2\cos\sigma)/\zeta_1 D_1\}]} \{r_0^{-1}[ir_0^{-1}(1 - e^{ir_0 t}) - t] \tag{20}$$

$$- \frac{1}{\zeta[r_0 + \{(\lambda_3 - eHl^2\cos\sigma)/\zeta_1 D_1\}] + r_0} [\frac{i(e^{it\{\zeta[r_0 + \{(\lambda_3 - eHl^2\cos\sigma)/\zeta_1 D_1\}] + r_0\}} - 1)}{\zeta[r_0 + \{(\lambda_3 - eHl^2\cos\sigma)/\zeta_1 D_1\}] + r_0} - t]\},$$

$$I_{5\omega}(t) = \frac{-i\tilde{\Pi}_{10}}{r_0 - \zeta[r_0 + \{(\lambda_3 - eHl^2\cos\sigma)/\zeta_1 D_1\}]} \left\{ \frac{i[1 - e^{it\{r_0 - \zeta[r_0 + \{(\lambda_3 - eHl^2\cos\sigma)/\zeta_1 D_1\}]\}}]}{r_0 - \zeta[r_0 + \{(\lambda_3 - eHl^2\cos\sigma)/\zeta_1 D_1\}]} - t \right\}$$

$$- \frac{\tilde{A}}{\zeta[r_0 + \{(\lambda_3 - eHl^2\cos\sigma)/\zeta_1 D_1\}]} \{r_0^{-1}[ir_0^{-1}(1 - e^{ir_0 t}) - t] \tag{21}$$

$$- \frac{1}{r_0 - \zeta[r_0 + \{(\lambda_3 - eHl^2\cos\sigma)/\zeta_1 D_1\}]} [\frac{i(1 - e^{it\{r_0 - \zeta[r_0 + \{(\lambda_3 - eHl^2\cos\sigma)/\zeta_1 D_1\}]\}})}{r_0 - \zeta[r_0 + \{(\lambda_3 - eHl^2\cos\sigma)/\zeta_1 D_1\}]} - t]\}.$$

### D. Transverse and axial displacement solutions

In order to obtain significant insights into the body's position in space during maneuvers near other bodies, examinations have been conducted on its transverse and axial displacement. By integrating equation (13) concerning time, the equation for transverse displacements can be derived, signifying

$$\eta(t) = \eta(0) + \int_0^t [\omega(0) + im^{-1}e^{i\theta_0}\{r_0^{-1}f(1 - e^{ir_0\tau}) - f_3[\Gamma_0 t + I_\omega(\tau)]\}]d\tau$$

$$= \eta(0) + [\omega(0) + i(mr_0)^{-1}e^{i\theta_0}f]t + m^{-1}e^{i\theta_0}\{r_0^{-2}f(e^{ir_0 t} - 1) - if_3[(\Gamma_0 t^2/2) + I_\eta(t)]\}, \tag{22}$$

where

$$I_\eta(t) = [(\sqrt{\zeta_1} + \sqrt{\zeta_2})I_{1\eta}(t) + (\sqrt{\zeta_1} - \sqrt{\zeta_2})I_{2\eta}(t)]/2\zeta, \tag{23}$$

$$I_{1\eta}(t) = \int_0^t I_{1\omega}(\tau)d\tau$$

$$= \frac{-i\Pi_{10}}{\zeta[r_0 + \{(\lambda_3 - eHl^2\cos\sigma)/\zeta_1 D_1\}] + r_0} \{ \frac{i}{\zeta[r_0 + \{(\lambda_3 - eHl^2\cos\sigma)/\zeta_1 D_1\}] + r_0}$$

$$\times [t + \frac{i(e^{it(\zeta[r_0 + \{(\lambda_3 - eHl^2\cos\sigma)/\zeta_1 D_1\}] + r_0)} - 1)}{\zeta[r_0 + \{(\lambda_3 - eHl^2\cos\sigma)/\zeta_1 D_1\}] + r_0}] - \frac{t^2}{2}\} + \frac{A}{\zeta[r_0 + \{(\lambda_3 - eHl^2\cos\sigma)/\zeta_1 D_1\}]} \tag{24}$$

$$\times \{r_0^{-1}[ir_0^{-1}\{t - ir_0^{-1}(1 - e^{ir_0 t})\} - \frac{t^2}{2}] - \frac{1}{\zeta[r_0 + \{(\lambda_3 - eHl^2\cos\sigma)/\zeta_1 D_1\}] + r_0}[-\frac{t^2}{2}$$

$$+ \frac{i}{\zeta[r_0 + \{(\lambda_3 - eHl^2\cos\sigma)/\zeta_1 D_1\}] + r_0}[t + \frac{i(e^{it(\zeta[r_0 + \{(\lambda_3 - eHl^2\cos\sigma)/\zeta_1 D_1\}] + r_0)} - 1)}{\zeta[r_0 + \{(\lambda_3 - eHl^2\cos\sigma)/\zeta_1 D_1\}] + r_0}]\},$$

$$I_{2\eta}(t) = \int_0^t I_{2\omega}(\tau)d\tau$$

$$= \frac{-i\tilde{\Pi}_{10}}{r_0 - \zeta[r_0 + \{(\lambda_3 - eHl^2\cos\sigma)/\zeta_1 D_1\}]} \{ \frac{i}{r_0 - \zeta[r_0 + \{(\lambda_3 - eHl^2\cos\sigma)/\zeta_1 D_1\}]}$$

$$\times [t + \frac{i(1 - e^{it(r_0 - \zeta[r_0 + \{(\lambda_3 - eHl^2\cos\sigma)/\zeta_1 D_1\}])} - 1)}{r_0 - \zeta[r_0 + \{(\lambda_3 - eHl^2\cos\sigma)/\zeta_1 D_1\}]}] - \frac{t^2}{2}\} - \frac{\tilde{A}}{\zeta[r_0 + \{(\lambda_3 - eHl^2\cos\sigma)/\zeta_1 D_1\}]} \tag{25}$$

$$\times \{r_0^{-1}[ir_0^{-1}\{t - ir_0^{-1}(1 - e^{ir_0 t})\} - \frac{t^2}{2}] - \frac{1}{r_0 - \zeta[r_0 + \{(\lambda_3 - eHl^2\cos\sigma)/\zeta_1 D_1\}]}[-\frac{t^2}{2}$$

$$+ \frac{i}{r_0 - \zeta[r_0 + \{(\lambda_3 - eHl^2\cos\sigma)/\zeta_1 D_1\}]}[t + \frac{i(1 - e^{it(r_0 - \zeta[r_0 + \{(\lambda_3 - eHl^2\cos\sigma)/\zeta_1 D_1\}])} - 1)}{r_0 - \zeta[r_0 + \{(\lambda_3 - eHl^2\cos\sigma)/\zeta_1 D_1\}]}]\}.$$

In a manner akin to the scenario involving transverse displacements, the axial displacement can be determined by integrating equation (17) to get

$$\eta_3(t) = \eta_3(0) + \omega_3(0)t + \frac{f_3}{2m}t^2 + \frac{i}{2m}(\tilde{f} I_{3\eta} - f \tilde{I}_{3\eta}), \tag{26}$$

where

$$I_{3\eta}(t) = \int_0^t [-\frac{i\Gamma_0}{r_0}(1 - e^{-ir_0\tau}) + I_{3\omega}(t)]d\tau$$
$$= \frac{i\Gamma_0}{r_0}[\frac{i(e^{-ir_0 t} - 1)}{r_0} - t] + \zeta_3 I_{4\eta}(t) + \zeta_4 I_{5\eta}(t), \tag{27}$$

$$I_{4\eta}(t) = \frac{-i\Pi_{10}}{\zeta[r_0 + \{(\lambda_3 - eHl^2\cos\sigma)/\zeta_1 D_1\}] + r_0} \{\frac{i}{\zeta[r_0 + \{(\lambda_3 - eHl^2\cos\sigma)/\zeta_1 D_1\}] + r_0}$$
$$\times [t + \frac{i(e^{it\{\zeta[r_0 + \{(\lambda_3 - eHl^2\cos\sigma)/\zeta_1 D_1\}] + r_0\}} - 1]}{\zeta[r_0 + \{(\lambda_3 - eHl^2\cos\sigma)/\zeta_1 D_1\}] + r_0}] - \frac{t^2}{2}\} + \frac{A}{\zeta[r_0 + \{(\lambda_3 - eHl^2\cos\sigma)/\zeta_1 D_1\}]}$$
$$\times \{r_0^{-1}[ir_0^{-1}\{t - ir_0^{-1}(1 - e^{ir_0 t})\} - \frac{t^2}{2}] - \frac{1}{\zeta[r_0 + \{(\lambda_3 - eHl^2\cos\sigma)/\zeta_1 D_1\}] + r_0}[-\frac{t^2}{2}$$
$$+ \frac{i}{\zeta[r_0 + \{(\lambda_3 - eHl^2\cos\sigma)/\zeta_1 D_1\}] + r_0}[t + \frac{i(e^{it\{\zeta[r_0 + \{(\lambda_3 - eHl^2\cos\sigma)/\zeta_1 D_1\}] + r_0\}} - 1)}{\zeta[r_0 + \{(\lambda_3 - eHl^2\cos\sigma)/\zeta_1 D_1\}] + r_0}]\}, \tag{28}$$

$$I_{5\eta}(t) = \frac{-i\Pi_{10}}{r_0 - \zeta[r_0 + \{(\lambda_3 - eHl^2\cos\sigma)/\zeta_1 D_1\}]} \{\frac{i}{r_0 - \zeta[r_0 + \{(\lambda_3 - eHl^2\cos\sigma)/\zeta_1 D_1\}]}$$
$$\times [t + \frac{i(1 - e^{it\{r_0 - \zeta[r_0 + \{(\lambda_3 - eHl^2\cos\sigma)/\zeta_1 D_1\}]\}}}{r_0 - \zeta[r_0 + \{(\lambda_3 - eHl^2\cos\sigma)/\zeta_1 D_1\}]}] - \frac{t^2}{2}\} - \frac{\tilde{A}}{\zeta[r_0 + \{(\lambda_3 - eHl^2\cos\sigma)/\zeta_1 D_1\}]}$$
$$\times \{r_0^{-1}[ir_0^{-1}\{t - ir_0^{-1}(1 - e^{ir_0 t})\} - \frac{t^2}{2}] - \frac{1}{r_0 - \zeta[r_0 + \{(\lambda_3 - eHl^2\cos\sigma)/\zeta_1 D_1\}]}[-\frac{t^2}{2}$$
$$+ \frac{i}{r_0 - \zeta[r_0 + \{(\lambda_3 - eHl^2\cos\sigma)/\zeta_1 D_1\}]}(t + \frac{i(1 - e^{it\{r_0 - \zeta[r_0 + \{(\lambda_3 - eHl^2\cos\sigma)/\zeta_1 D_1\}]\}}}{r_0 - \zeta[r_0 + \{(\lambda_3 - eHl^2\cos\sigma)/\zeta_1 D_1\}]})\}, \tag{29}$$

## IV. CONCLUSION

The present study has been focused on achieving the novel analytic solution for the motion of a spinning axisymmetric charged rigid body. Therefore, Euler's equations have been utilized to derive the equations of motion that governs the rigid body's rotatory motion. Due to the absence of torque along the spin axis and the rigid body possesses an almost axisymmetric characteristic, the spin rate remains relatively stable. The analytical solutions of the rigid body for the attitude, rotational, and translational motion have been derived through an assumption of small angular deviations of the spin axis with respect to a fixed direction in space. These solutions encompass various parameters such as angular velocities, Euler's angles, transverse velocities, transverse displacements, axial velocity, and axial displacement.